\documentclass[%
 reprint,
superscriptaddress,
frontmatterverbose, 
 amsmath,amssymb,
 aps,
prstper,
]{revtex4-2}

\usepackage{graphicx}% Include figure files
\usepackage{dcolumn}% Align table columns on decimal point
\usepackage{bm}% bold math
\usepackage{svg}
\usepackage{booktabs}
\usepackage{tabularx}
\usepackage{subcaption}

\usepackage[breaklinks]{hyperref}  
\hypersetup{colorlinks=true,urlcolor=blue,citecolor=blue,linkcolor=blue}   
\usepackage{enumitem}          % package for handling list formatting
\setlist{nosep}                 % Tightest spacing for lists. `noitemsep` is more relaxed

\PassOptionsToPackage{hyphens}{url}\usepackage{hyperref}
\usepackage{url}

  {\list{}{\leftmargin=0.3in\rightmargin=0.3in}\item[]}%
  {\endlist}

\begin{document}

\preprint{APS/123-QED}

\title{Investigating Student Participation in Quantum Workforce Initiatives}% Force line breaks with \\
%%\thanks{A footnote to the article title}%

\author{Michael B. Bennett}%
\affiliation{Q-SEnSE NSF Quantum Leap Challenge Institute, Boulder CO USA 80309}
%\affiliation{University of Colorado Boulder, Boulder CO USA 80309}

\author{Joan \'Etude Arrow}
\affiliation{Q-SEnSE NSF Quantum Leap Challenge Institute, Boulder CO USA 80309}
%\affiliation{University of Colorado Boulder, Boulder CO USA 80309}

\author{Sasha Novack}
\affiliation{ATLAS Institute, University of Colorado Boulder, Boulder CO USA 80309}

\author{Noah D. Finkelstein}
\affiliation{Department of Physics, University of Colorado Boulder, Boulder CO USA 80309}

%%\collaboration{MUSO Collaboration}%\noaffiliation

%%\date{\today}% It is always \today, today,

\begin{abstract}
As the focus of quantum science shifts from basic research to development and implementation of applied quantum technology, calls for a robust, diverse quantum workforce have increased.  However, little research has been done on the design and impact on participants of workforce preparation efforts outside of R1 contexts.  In order to begin to answer the question of how program design can or should attend to the needs and interests of diverse groups of students, we performed interviews with students from two Colorado-based quantum education/workforce development programs, one in an undergraduate R1 setting and one in a distributed community setting and serving students largely from two-year colleges.  Through analysis of these interviews, we were able to highlight differences between the student populations in the two programs in terms of participation goals, prior and general awareness of quantum science, and career interest and framing of career trajectories.  While both groups of students reported benefits from program participation, we highlight the ways in which students' different needs and contexts have informed divergent development of the two programs, framing contextual design of quantum education and workforce efforts as an issue of equity and representation for the burgeoning quantum workforce.

\end{abstract}

\maketitle

%\tableofcontents

%%%%%%%%%%%%%%%%%%%%%%%%%%%%%%%%%%%%%%%%%%%%%%%%%%%%
\section{Introduction}
\label{sec:intro} 

As the ``Second Quantum Revolution'' picks up steam, the focus of quantum science is broadening from a focus on understanding the physical ``rules'' of quantum systems to manipulating those systems for practical purposes, such as communication.  Motivations for quantum science are also expanding from basic, curiosity-driven research to include practical and product-focused development and commercialization \cite{jaeger2018second,deutsch2020harnessing,dowling2003quantum}.  In the United States, the promise of quantum computing and communication technology led to the 2018 passing of the National Quantum Initiative (NQI) Act and the creation of the National Quantum Coordination Office (NQCO), the Quantum Economic Development Consortium (QED-C), and other bodies aimed at ensuring the creation of a robust quantum ecosystem within the U.S to support both societal and State activities \cite{NQI_Act,raymer2019us}.

As the shift from the laboratory to the production floor accelerates, institutions of higher education are accelerating their own efforts to design and implement education and workforce training initiatives that meet the increasing industry demand for quantum-trained workers.  Recent foundational STEM education research has demonstrated the broad need of the burgeoning quantum industry for workers \cite{hughes_assessing_2022, fox_preparing_2020}, as well as documenting efforts of higher education to meet those needs \cite{cervantes2021overview,aiello_achieving_2021,perron_quantum_2021,meyer_todays_2022, asfaw2022building}.  Interestingly, multiple investigations have revealed industry's increasing demand for workers, at all levels, demonstrating competency with technical and analytical skills but \textit{not} necessarily with quantum science specialist training \cite{fox_preparing_2020,hasanovic_quantum_2022,qedc_tech_report}.  Much of the discussion surrounding quantum education, including at the federal level, has thus focused on the importance of developing training opportunities for these so-called ``quantum-aware'' and ``quantum-conversant'' workers.  The 2022 Subcommittee on Quantum Information Science (SCQIS) \textit{Workforce Development National Strategic Plan} describes the situation: \textit{``Only about half of the roles sought by industry require QIST proficiency. The remainder rely on workers with, at most, a basic awareness of QIST. The desired education levels span bachelor’s, master’s, and doctoral degree recipients.''} \cite{SCQIS_report}

Despite the apparent need for non-specialist quantum training initiatives in higher education, much of the research to-date in quantum education has highlighted degree programs and workforce development initiatives at Carnegie-classified R1 universities \cite{center2018carnegie}.  Of course, given the early state of the quantum education research enterprise and the fact that R1 universities are by definition more likely to produce and house the researchers -- and students -- participating in quantum education, this is perhaps to be expected.  Still, however, we note that such universities may focus primarily on quantum specialist training in their workforce development efforts, even at the undergraduate level, rather than the type of ``quantum-conversant'' training needed by the workers described above.  Jobs requiring this type of training may include engineering or technician roles such as optics technician, programmer, electrical engineer, etc., and are more likely to be jobs where \textit{in situ} training can be appropriate for training workers via onboarding, internships, apprenticeships, and other experiential methods.  

\subsubsection{Representation and the Student Experience in Quantum Science}

Encouragingly, the QIS discourse has, even at the top level, articulated an explicit need not only for the creation of a quantum workforce, but for a diverse, representative workforce.  Both the SCQIS Strategic Report quoted above and the 2023 NQI Advisory Committee (NQIAC) \textit{Report on Renewing the National Quantum Initiative} \cite{NQIAC_report} note the importance of a broadly inclusive QIS education enterprise.  From the NQIAC report:   \textit{``Efforts targeting all education levels, but particularly K-12 and community colleges, are necessary to make sure the future QIST workforce is more diverse and inclusive than the workforce of today.''}

2021 data from the \textit{Integrated Post-Secondary Education Data System} (IPEDS) suggests that students typically underrepresented in four-year colleges(Black, Hispanic, Native) are greatly over-represented at the two-year college level (39.44\% enrollment vs 31.5\% census data) \cite{IPEDS,census_data}, yet these institutions to-date are deeply underresourced \textit{and} under-studied in the quantum education space.  We also note that the majority of research on quantum education and workforce development has focused either on educator efforts or on the needs of industry, with few studies investigating the student perspective (understanding of QIS, disposition toward careers, etc.) \cite{rosenberg2024science}.  Questions of student interest, disposition, and affect also call to mind the importance of non-formal experiences, which can span contexts and venues and often focus explicitly on developing affect in participants.  Researchers and educators alike must attend to all of these considerations if we are to successfully entice an entire workforce's worth of bright young minds.

\subsubsection{Purpose of This Study}
\label{sec:intro-purpose}

We thus identify as a crucial topic for study the student experience in quantum education and workforce development initiatives, and articulate three questions that form the basis for this study.

\begin{itemize}
    \item How do students participating in quantum education programs, both in R1 settings and in ``non-traditional'' settings, describe their path to participation in QIS as a potential career field, and how do they describe the impacts of program participation on their quantum career interests?
    
    \item What factors in program design can support the participation of these diverse groups of students in quantum education and workforce activity?
    
    \item How do these students participating in quantum education programs describe their objectives both for program participation and for their future career activity?
\end{itemize}

This study begins to address these questions through a qualitative investigation of students in two Colorado-based quantum education and workforce programs, one implemented at a large R1 university, and one implemented at several smaller schools, most of which are two-year community colleges.  The development and analysis of student profiles based on interviews with program participants allows for the comparison and contrast of student goals, interests, and initiatives in these two populations; we discuss as well the implications for program design, ways in which the two programs studied can adapt and have adapted to meet the needs of their respective student populations, and generalizable lessons for broader, diversity-focused quantum education program design.

%%Since the National Quantum Initiative (NQI) Act was passed in 2018, the need to develop a workforce to support the emerging quantum industry has become apparent. Past research has focused on the needs of the quantum industry, the skills that these companies are looking for, and the educational programs that can prepare a workforce with the skills to meet these needs \cite{fox_preparing_2020} \cite{hughes_assessing_2022} \cite{aiello_achieving_2021} \cite{hasanovic_quantum_2022} \cite{perron_quantum_2021} \cite{meyer_todays_2022}. In this work, we explore the perspectives of students who have participated in quantum workforce development initiatives at Q-SEnSE to better understand the needs and goals of students considering entering the quantum workforce. We interview students who participated in either the Quantum Research Exchange (QRX) program which focuses on students at 2-year institutions in the Denver-Boulder area, and undergraduates at CU Boulder who participated in the Quantum Scholars program. 

%%%%%%%%%%%%%%%%%%%%%%%%%%%%%%%%%%%%%%%%%%%%%%%%%%%%
\section{Study Context, Background, \& Positionality}
\label{sec:context}

This study takes place in the context of Colorado's rich quantum ecosystem.  Two of this manuscript's authors are affiliated with the Q-SEnSE NSF Quantum Leap Challenge Institute (QLCI) \cite{qsense_website}, which is headquartered at the University of Colorado Boulder (CU Boulder) \cite{cu_boulder_website}; the other two authors are affiliated with the University's Department of Physics and the Department's Physics Education Research Group \cite{cu_physics_website,cuper_website}.  CU Boulder itself enjoys robust connection with a large (and increasing) number of quantum industry players in the Denver-Boulder area east of the Colorado Rockies.  Indeed, both programs investigated in this study benefited from participation from numerous industry partners, which is a key and relatively unique element of their implementation specifically in Colorado.  In this section we describe the context in which the study takes place: the institutional background supporting the development of both programs; their design and implementation, including target participant populations; and the positionality and demographics of both the research team members and the student populations from which study participants were drawn.

\subsection{Institutional Background}
\label{sec:context-institution}

As mentioned above, this study focuses on two programs and compares the experiences of a small population of students participation in each program.  One of these programs is implemented at the CU Boulder campus and supported by the physics department and the College of Engineering and Applied Sciences; the other is implemented throughout the Denver-Boulder area and is sponsored by the Q-SEnSE NSF Quantum Leap Challenge Institute.  We here describe the two institutional contexts in which these programs exist respectively.

\subsubsection{Q-SEnSE and the QLCI Ecosystem}
\label{sec:context-qlci}

In 2019, the National Science Foundation introduced the Quantum Leap Challenge Institute (QLCI) program as part of its Quantum Leap Big Idea \cite{durakiewicz2018enabling,BigIdeaQLwebsite}.  Three QLCIs were funded in the inagural round of awards in 2020, with two more funded the following year.  As one of those five QLCIs, Q-SEnSE is a national leader in quantum sensing and metrology research, drawing on a strong tradition of both at its lead institution, CU Boulder, as well as at ten other academic institutions.  As part of its award, Q-SEnSE funds a Director for Education and Workforce Development whose role is to set educational strategy for the center, design and implement programs, conduct assessments, etc.  The Director (this manuscript's first author) is housed at CU Boulder and, as a consequence, a majority of Q-SEnSE's award-funded educational initiatives are implemented in the Boulder-Denver area, despite the fact that Q-SEnSE is technically a distributed center.  All Q-SEnSE member institutions are either R1 Research Universities or national labs.  Indeed, all five QLCIs are headquartered at large, top-15 R1 institutions \cite{USNewswebsite}; we note these classifications given that, altogether, the QLCI program represents so far a \$125 million investment by the federal government.  A key motivating factor in Q-SEnSE's educational strategy is awareness of this ecosystem and a desire to ensure the participation of non-R1 stakeholders, such as those at community colleges, in quantum education and workforce development.

Q-SEnSE's Director of Education and Workforce Development is in large part the sole education and workforce authority for the center and, as a consequence, is able to make strategic decisions for the center's education efforts with relative agility and autonomy.  This autonomy also means that center capital can be brought to bear to meet strategic education goals quickly and efficiently; by the same token, however, it also means that, similar to many other grant-funded education initiatives, Q-SEnSE's education and workforce efforts are somewhat decoupled and isolated from the center's main scientific activity.  As a consequence, Q-SEnSE often relies on external partners for the ``personpower'' required to turn strategic designs into successful implementations.  In such partnerships, Q-SEnSE tends to serve largely as the ``administrative center'' of efforts, coordinating, funding, and leading implementations that are often more directly overseen by partners from the local context rather than by Q-SEnSE leaders.  The motivations and rationale for this paradigm beyond administrative necessity will be detailed in Section \ref{sec:context-qrx} alongside a description of the Q-SEnSE sponsored program that forms the basis for this study.

\subsubsection{CU Boulder Physics Department}
CU Boulder itself is an international leader in physics, particularly within quantum sciences and quantum education. CU Boulder Physics is among the largest programs in the United States —-with the fourth largest undergraduate (Bachelor’s) degree program and largest doctoral (Ph.D.) program.  While there are many nationally leading research fields within the department, efforts within quantum physics and quantum education have been areas of long-standing investment and attention.  Besides headquartering Q-SEnSE, the campus also boasts several quantum centers (an NSF Physics Frontier Center, a Department of Energy Quantum Engineering Initiative, and multiple NSF Science Technology Centers ) all brought together through the CUbit Quantum Initiative.  Four Nobel Prizes, several MacArthur Fellows, and numerous other awards have been given to the CU Physics faculty for their work in advancing QISE.

Efforts focussed on understanding and advancing quantum education date back to the early 2000’s (and have prior antecedents) that span from the Physics 2000 website, the PhET interactive simulation project \cite{perkins2006phet}, and the physics education research (PER) group that was formed in 2003.  The PhET sims (including work on making quantum science accessible and understandable) have achieved over 1.5B downloads. A mainstay of research in the PER group has been studies of student understanding of quantum concepts, student reasoning around these ideas, student interpretation of quantum phenomena, and associated curricula to advance student access to and understanding of quantum physics \cite{mckagan2010design,sadaghiani2015quantum,corsiglia2023intuition,borish2023seeing,meyer2022investigating} These efforts range from pre-college to graduate level engagement and span both focus on both theoretical and experimental courses/ environments. 

\subsection{Program Context}
This study investigates two programs: the Q-SEnSE Quantum Research Exchange and the CU Boulder Physics Department's Quantum Scholars.  We here briefly describe each program in terms of their design philosophy, objectives, and implementation.

\subsubsection{The Quantum Research Exchange}
\label{sec:context-qrx}

The Quantum Research Exchange (QRX) is an ``internship readiness'' program focusing on providing students, primarily those not at R1s and those from backgrounds historically underrepresented in physics, with opportunities to explore career pathways in the quantum sector, build professional skills, and gain confidence in their ability to fully participate in QIS.  Building on the program designer's expertise with affect-focused education design and partnership-based program development, particularly in the space outside of classrooms \cite{gutierrez_developing_2008,cole_fifth_2006}, and informed by both the Design-Based Implementation Research paradigm and the Community of Practice model \cite{fishman2013design,cox2005communities,li2009evolution}, QRX brings students together in yearly cohorts whose members support one another through both professional development activity and foundational quantum content exploration.

Students are recruited from a variety of schools in the Denver-Boulder area, usually two-year schools in the Colorado Community College System (CCCS).  The program partners with STEM faculty members, who serve as leads and liaisons at their school; QRX works with each partner school to tailor recruitment, program scheduling, and offerings to the unique context and needs of partners' students, while also offering a centralized curriculum of largely-online events and opportunities shared across the entire program.  Program-wide offerings include resum\'e-building workshops, interview and networking practice, guest lectures from industry members, industry networking opportunities, ``Quantum 101'' lectures presenting quantum concepts at a basic conversational level, quantum ``hackathon'' events, and more.  Overall, the programming aims to facilitate the development of students' sense of belonging as a group of like-minded, savvy, young professionals in the quantum space.  Industry partners play a key role in the program's success, and participate in the program's online community while offering opportunities for connection and even employment.

\subsubsection{CU Quantum Scholars}
\label{sec:context-qscholars}

The Quantum Scholars program launched in spring of 2023 to provide funding and opportunities to help students learn about the quantum field, including connecting students with local industry and quantum technology. While CU was already recognized as a national leader in quantum physics and engineering, the new scholarship program was designed help grow, broaden, highlight, diversify and advance efforts in quantum education and workforce development. The quantum scholars program, initiated through joint support of the Department of Physics and College of Engineering and Applied Sciences, supports undergraduates in physics, engineering and computer science with the aim to advance quantum education and workforce development through professional development, co-curricular activities, and industrial engagement \cite{qscholars_website}.  The program was initially modeled on the successful CU Boulder Kiewit Design-Build Scholars Program, a year-long program that engages engineering students in meaningful conversations about future careers \cite{kiewit_website}.

The inaugural cohort of undergraduates across quantum- related fields began with 53 students, 21 of whom received \$2500 in fellowship funding. Of the scholars, 36\% identified as non-male, and 38\% as people of color (Hispanic/Latinx, Black, Native American, AAPI, Mixed Race). Student motivation for participating in the program (as determined from their applications), ranged from finding a bridge among traditional disciplines (chemistry, physics and CS) and learning about what quantum sciences looks like beyond the classroom, to building and participating a community that supports the diverse array of learners at CU (that historically have not been represented in QISE fields).  Others recognized the need for and impact of funding to complement their interests, enabling students to participate in professional development in quantum fields without taking an additional job.

The cohort of students met with faculty advisors and a graduate student program manager once to twice per month over the academic term.  Quantum Scholars activities ranged from professional presentations on the leading scholarship in quantum research, to the practice of building a company in QISE fields, to tours of the national labs (NIST) and regional companies.  Additional activities included professional development (preparing for an internship, applying to graduate school in quantum fields), creating and staffing a quantum career fair, and social enterprises (ranging from creating movie nights to a slack channel, and undergraduate club) focussed to quantum. Since inception the program has grown (now featuring 75+ undergraduates with many student repeating), more funding (40+ fellowships), more programs (including a hackathon), and student-led / designed activities.

\subsection{Demographics}
\label{sec:context-demo}

As will be described below, we originally had intended to survey students from both programs in order to obtain a representative sample of demographics; unfortunately, low survey response meant that we were unable to obtain demographic information for our sample set of subjects that could be published without running afoul of low statistics concerns (representativeness, likelikhood of publishing demographic information that could be traced to a particular individual, etc.).  However, we here describe in broad terms the demographics of students attending the schools supported by the two programs described above.

CU Boulder student demographic information is published annually by the CU Office of Data and Analytics and includes information on new, continuing, and transfer undergraduates as well as new and continuing graduate students.  These data include breakdowns by college/division as well as major.  In the Fall 2023 semester, CU Boulder reported 485 undergraduate physics majors enrolled in either the Physics or Engineering Physics major, 435 of which were domestic (CU Boulder collapses ``international'' students into a single race/ethnicity category).  Similar to the school as a whole, the physics department population overrepresents white and asian students (65\% and 11.3\% respectively for the department, compared to 59.3\% and 5.9\% respectively for the US \cite{CU_stats,census_data}) and underrepresents other races/ethnicities (13.5\% combined for Black, Native American/Alaskan, Hispanic/Latino, and Native Hawaiian/Pacific Islander, compared to 38.3\% nationally).  Quantum Scholars recruits students starting from second year in Physics, Computer Science, and Engineering majors; demographics for the program as a whole are reported above with the program description -- we note that its URM statistics more closely resemble national statistics.

Demographics for QRX students are slightly more difficult to estimate given that students come from a variety of schools and the individual departments of each school are much smaller than CU Boulder's physics department.  CU Denver, for example, reported 53 physics majors at the end of its Fall 2023 term, 38\% of which were ``minority'' (non-white) students \cite{CUDenver_stats}.  The Colorado Community College System does not publish complete demographic information of its students or break statistics down by major, but it does report proportions of ``students of color'' enrolled at each school; for the three CCCS schools participating in QRX, those proportions were 62.10\%, 60.10\%, and 35.10\% for the 2022-2023 academic year \cite{CCCS_stats}.

We report these demographic estimates not to position one or the other of the two programs studied here as ``better'' as a function of the populations they purport to serve; naturally, the proof of a beneficial program is in the assessment.  Rather, we highlight that the two programs are, likely serving very different sets of students -- a point which will become increasingly pertinent as we discuss some of the findings regarding program design and benefits in Section \ref{sec:discussion}.

%%\subsubsection{Research Positionalities}

%%As a brief aside before attending to study design and methods, we note here the positionalities of the paper's authors.  Both identify as white; one is a cis man and one is a trans woman.  Bennett attended a four-year undergraduate liberal arts college and worked at a museum for two years before completing graduate study at a top-ten R1 for his field of study.  Arrow [Joan fills out 1-2 sentences max.]

%%%%%%%%%%%%%%%%%%%%%%%%%%%%%%%%%%%%%%%%%%%%%%%%%%%%
\section{Methodology}
\label{sec:methodology}

This project's main goal was to probe the attitudes, beliefs, and motivations of students participating in the two programs described above, and to generate a basic understanding of the ways in which programs designed for different groups of stakeholders might meet the needs of their target populations.  Data was taken in the Spring 2023 semester and the Spring 2024 semester for the Quantum Scholars and QRX students, respectively (at the end of each program's first year of implementation).  We originally designed the study with a wide administration of surveys to create a broad picture of program participation with both quantitative and qualitative elements; interview subjects were originally to be recruited from the group of students who completed a post-participation survey of their program and interview data would focus on interesting elements of individual students' participation in their particular program.  However, for both Quantum Scholars and QRX, we struggled to achieve a strong survey completion rate.  In particular, the distributed nature of the QRX program made it difficult to encourage students directly to fill out the survey as their program participant came to an end; students at different schools ended the spring semester at different times, with different post-semester constraints and obligations.  Instead, we ended up relying more on the interview data from both programs, producing a smaller but richer data set.

As a consequence, we note that the study as conducted likely has limited generalizability compared to the study as originally intended; without a high degree of statistical precision from surveys, it is difficult to compare, for example, increases in quantum awareness between the two programs, or between current and future implementations of either program.  Nevertheless, the interview data we collected did form a relatively consistent picture of both the students participating in the two programs and the nature of participation, with stark qualitative differences presenting themselves relatively quickly during data collection and analysis.  The data for this study, then, should be looked at as ``thematic snapshots'' of two quantum-focused enrichment programs in the Colorado quantum ecosystem.

\subsection{Interviews}
\label{sec:methodology-interviews}

As mentioned above, we used a semi-structured interview protocol that was the same for both programs apart from labels and names of programs, schools, etc.  The interview protocol was designed with three major data collection goals in mind: (1) to ascertain exactly when, where, and how the subject student became aware of QIS as a field and the student's path to program participation; (2) solicit the student's understanding of QIS as a career field and their expectations for participation in that field, especially via program participation; and (3) probe the shape of their general STEM identity and their sense of belonging in STEM generally and QIS specifically.  Students were informed at the start of the interview about the nature of the study and told that they could stop the interview at any time and that all questions were optional and would not affect their participation in their program in any capacity.  After a set of ``icebreaker'' questions designed to help the subject feel comfortable in the interview space, the interviewer steered the conversation into each of the three topic areas above while leaving space for subject-led digressions.  At the end of the interview, subjects were given the opportunity to share any lingering thoughts or direct the conversation as they chose.

In total, 12 interviews were taken: six from CU Boulder's Quantum Scholars in spring 2023 and six from Q-SEnSE's QRX program in spring 2024.  Each set of interviews were conducted with the first cohort of students to participate in the program.  As mentioned above, interviews were conducted via Zoom, largely for convenience since students were not all attending the same school.  Interviews were recorded onto a research computer and the audio was transcribed using a two-pass format: voice recognition software was used to create a base transcript, then the researchers passed through the transcript while listening to the audio and performing spot checks to correct any errors produced by the software.  Interview transcripts and their associated audio files were then imported into MaxQDA for coding.  Interview subjects were given pseudoynms prior to analysis.

\subsection{Coding Scheme and Analysis of Interviews}
\label{sec:methodology-coding}

The codebook for analysis was produced following the collection of the Quantum Scholars interviews but before the collection of the QRX interviews.  In order to create the codebook, one of the investigators coded one of the Quantum Scholars interviews emergently, highlighting and assigning codes \textit{in vivo} based on subjects' responses to interview questions.  Following this, the investigator went through the emergent codes and grouped them together, combining where necessary to create a code system.  That system was then shared with another investigator without sharing the coded interview; the second investigator then coded the same interview and results were compared in order to determine ambiguities in the code system.  Once both investigators were in agreement on the definitions and usage of codes, they both independently coded half of a second interview, comparing their coding decisions and further refining the code bank.  Finally, the two investigators coded the second half of the second interview independently and compared their results for reliability and consistency (without making any modifications).  After the investigators were satisfied that the code system was robust, the rest of the data were coded independently using the system.

The code system ultimately consisted of four top-level codes, each with 3-5 subcodes.  The top level codes and their descriptions are shown in Table \ref{table:codes}.  Following the coding process, codes were parsed for patterns, common themes, similar answers, etc. among subjects, both within individual programs and between the two programs studied.  Those themes formed the basis of the analytical results presented in Section \ref{sec:results}.  

\begin{table}[]
    \centering
    \begin{tabularx}{\columnwidth}{r X}
        \hline \\
        \multicolumn{1}{c}{Top-Level Code}          &  \multicolumn{1}{c}{Code Description} \\
        \hline \hline \\
        Program-Related         &   The participant explicitly discusses aspects of the program in which they are involved.  Examples include: experiences in the program; feedback on program design; understanding of the program's goals; etc. \\
        Conception of QIS       &   The participation expresses some articulation of their own conception of QIS as a field.  Examples include: expressing an understanding of the field; articulating what sets QIS apart from other fields; describing activities that define QIS; etc. \\
        Conception of Physics   &   As with ``Conception of QIS'' but for physics broadly. Note that we did not articulate QIS as a ``sub-field'' of physics during interviews or during program participation.\\
        Career Pathways         &   The participant discusses their journey into participation in a career field (physics, QIS, etc.).  Examples include: motivation for attending college; post-degree intentions; career goals; etc. \\
        \hline \\
    \end{tabularx}
    \caption{The four top-level codes we used in analysis, as well as descriptions for each code based on the definitions we used for analysis.  These codes and their relevant subcodes were articulated emergently.}
    \label{table:codes}
\end{table}

\section{Data \& Results}
\label{sec:results}

\begin{figure*}
    \centering
    \includegraphics[trim={2.15cm 4.2cm 1.225cm 1.6cm}, clip, width=0.95\textwidth]{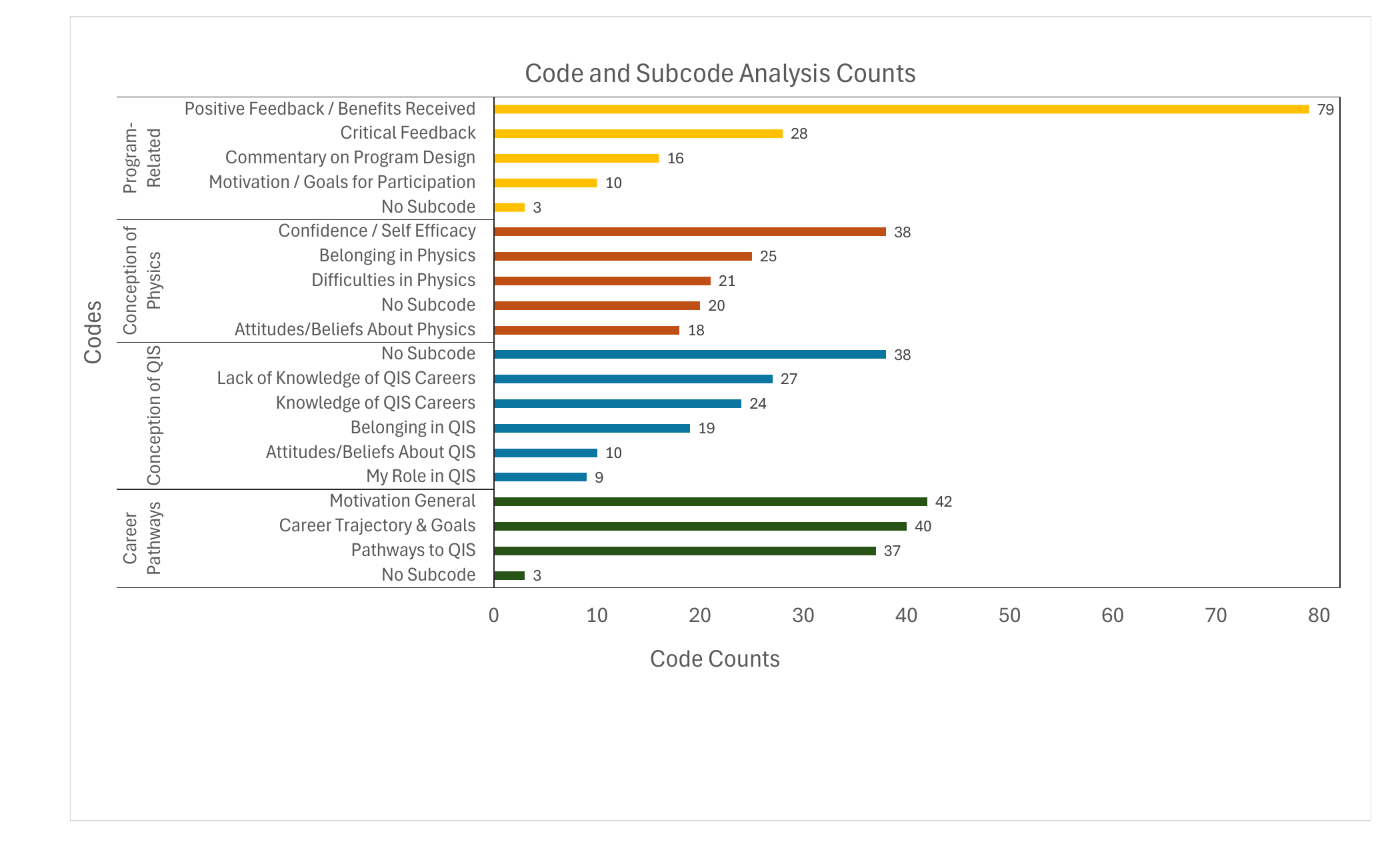}
    \caption{Code counts for each of the four top-level codes and their subcodes.  These totals represent coding assignments for all interview participants (i.e., they are not broken up by program).}
    \label{fig:codecounts}
\end{figure*}

A chart showing the number of counts for each code and subcode is presented as Figure \ref{fig:codecounts}.  In order to answer the questions posed in Section \ref{sec:intro-purpose}, we grouped our analysis around three major themes: 
\begin{enumerate}
    \item Program Participation (benefits received, critical program feedback, motivation for participation, etc.)
    \item Pathways to and Understanding of QIS (STEM background, awareness of quantum jobs/careers, etc.)
    \item Post-Participation Pathways and General STEM Affect (envisioned ongoing relationship to QIS, career objectives and goals, confidence and self-eficacy, etc.)
\end{enumerate}

For each of these themes, we will present results for each subgroup (Quantum Scholars participants and QRX participants), noting points of similarly and dissimilarity in preparation for interpretation in Section \ref{sec:discussion}.

\subsection{Program Participation}

For both Quantum Scholars and QRX, interview subjects expressed a generally positive outlook on their program participation.  Indeed, as shown in Figure \ref{fig:codecounts}, discussion of benefits accounted for a majority of program-related responses ($\approx 55\% $).  The types and experiences of those benefits, however, differed somewhat between the two programs.  Each of the six Quantum Scholars students expressed that their favorite ``single experience'' (a term multiple students used without being prompted) of the program was a tour of the nearby National Institute of Standards and Technology (NIST) laboratory; in terms of general program benefits, the students listed both the opportunity to learn more about the quantum industry through guest lectures and the opportunity to build community.  

In silhouette, the benefits reported by QRX students might appear similar: five of the six student expressed that they benefited greatly from participation in the end-of-program Quantum Recruiting Forum hosted by the Chicago Quantum Exchange (the sixth had not attended the event); multiple students also reported benefiting from opportunities to learn about the quantum industry from talks and networking events; and multiple students expressed affective benefits, either increased feeling of community in the QIS field or increased sense of self-efficacy.  However, the ways in which students talked about the benefits of their program experiences differed appreciably from the ways in which the Quantum Scholars students talked about their program benefits.  Across the board, QRX students couched their experience through the program in terms of career exploration, skill-building, and professional networking (indeed, when talking about community, multiple students explicitly touched on using professional networking site LinkedIn \cite{linkedin_site} to connect with other students and industry members).  By comparison, the Quantum Scholars group spoke of their program experiences mostly in terms of the academic or intellectual benefit enrichment they provided in parallel to the students' degree path and, when discussing the benefit of community, tended to speak more in terms of ``traditional'' collegiate experiences, such as finding new roommates, meeting new friends that persisted outside of the program, building clubs, etc.

Critical feedback on program participation and design also evidences distinct experiences between the two groups. Four of the six Quantum Scholars students reported that their least favorite element of program participation was program's focus on a generalized, industry-focused exploration of quantum science, rather than a more academic perspective on QIS.  Of the other two students, ``Jayden,'' an engineering student, expressed that the technical topics covered tended more toward ``physics topics'' (Jayden named trapped ions and photonics) and less toward ``engineering topics'' (superconductors), while ``Delaney,'' a physics major, expressed feeling as though she didn't have enough context to fully internalize the lectures.  For the QRX students, four of the six expressed a desire for more in-person events (as discussed in Section \ref{sec:context-qrx}, most events were online due to students' wide-ranging locations) as their strongest critical feedback.  Other critical feedback for QRX centered around a desire for more foundational quantum content and for professional development content more aware of students' level of prior professional awareness.

Taken together, these sets of feedback highlight a key divide in the interests, motivations, and goals for participation between the two groups, with the CU Boulder-based Quantum Scholars students framing their experience as more of an academic exercise and the broader, largely TYC-based QRX students framing program participation as a means of professional development.  We will dive more deeply into this dichotomy, and the nature of the communities build in each program, in Section \ref{sec:discussion}; however, it is worth noting here the distinction, since similar divergences show up when analyzing the other two major themes.

% harvey: nist tour, opportunities to learn more
% timothy: opportunities to learn, talks, etc., but single experience NIST
% jamie: NIST tour, lectures, but also feels like the community aspect is the most valuable aspect
% semaj: community was an important aspect, but single experience presentation from NIST, ``really connected with the quantum computing class i'm taking''
% jayden: community best aspect -- ``some of the speakers went over my head a little bit.'' other than that, the NIST tour
%delaney: different things than class: touring nist, visits, etc. ``gives you something to look forward too'' that isn't class

%% eli: hackathon, CQE event
%% mina: love the fact that we met a lot of people and networked with industry, in-person gatherings, community made me feel welcome, understanding of quantum careers
%% hillary: in-person events, talks from chris bishop et al. talks from industry helped me feel like this industry is something real for me to possibly participate in -- even if i wasn't going to grad school
%% eleanor: CQE event, networking, community. gained confidence in looking for jobs -- also touched on the name recognition of the program, sense of belonging in the field
%% bethany: connections to program personnel. ``it's actually shown me that i'm smart and i know what i'm doing,'' professional development
%% charity: expanded view of quantum, hackathon, networking and community, add things to resume, more welcomed in quantum

\subsection{Pathways to and Understanding of QIS}

Immediately upon analysis of participants' pathways to program participation and awareness of QIS specifically, we noticed an appreciable divergence between program participant groups.  Interview subjects were asked to articulate their current ``definitional'' understanding of QIS, recollect the first time they were made aware of QIS as a field, and connect the former to the latter in terms of experiences and influences that informed their understanding.  Without exception, QRX students articulated that participation in the program was their first foray into QIS as a field, if even they had encountered ``quantum'' as a concept before.  Two of six expressed that they had ``heard of quantum'' in high school; the other four expressed that recruitment drives for the QRX program were their first connection with quantum overall.  All six explicitly named a faculty member that had encouraged them to look into the program.

By comparison, participants in Quantum Scholars expressed broad and diverse pathways to quantum exposure.  Of the six interviewed, only one tied their first interaction with QIS to their recruitment to and participation in the Quantum Scholars program, concurrent with their enrollment in the CU Physics Department's modern physics course.  three other students expressed the opinion that their ``real'' introduction to quantum information science (not just ``quantum'') began in that same course or another course (e.g. quantum computing or quantum engineering), but all four expressed prior awareness of quantum concepts from as early an age as middle school.  One student reported that a prior internship at the Pacific Northwest National Laboratory exposed them to QIS; the sixth student reported that discussions with students at the end of high school and beginning of college had shaped their interest in QIS and led to them adding a quantum engineering minor to their degree path.  Together, the Quantum Scholars interviews revealed a level of prior exposure to quantum science not observed in the QRX student interviews.

When it came to articulating a strong understanding of QIS as a career field, however, differences in the program participant groups essentially vanished.  Even for those students in Quantum Scholars who expressed that they felt they understood QIS as a conceptual field of study, 12 of 13 students interviewed admitted that their understanding of QIS careers and jobs -- that is, what QIS professionals do on a day-to-day basis, job opportunities for entry into the field, long-term career prospects, etc. -- was appreciably underdeveloped, even after program participation.  When probed further, a number of students expressed feeling somewhat aware of the QIS career landscape, for example by tying their understanding to direct interactions they had in their program (e.g., through Quantum Scholars lectures from local QIS companies or from QRX networking events). One QRX student was able to articulate a fairly robust conception of job opportunities and entry-level ways to join the field, articulating an awareness of both technical needs in the field (``hardware and software stuff'') and increasing opportunities for non-technical workers (``\ldots people who take this technology to \ldots the consumer and train them on how to use it'').  This student, ``Eleanor,'' also articulated a basic awareness of the needs of ``incubator companies'' in the field raising money for research, as well as a surprisingly shrewd awareness of pathways to jobs in the field:

\textit{``I definitely am familiar with a lot more companies than I was before [participation in QRX], like companies that I hadn't even thought of before.  And I also have an idea of how to apply for these different jobs.  I think a lot of people are like, `oh, let me just go to Indeed or Glassdoor or something like that.  These jobs aren't on there.  You really need to go to the company website, or connect with people on LinkedIn, and I definitely have come to terms with that a lot more because of QRX.''}

We note that like the other QRX participants, her exposure to QIS as a field started with her participation in the program; she expressed as well that the program itself was largely responsible for her current understanding of QIS both as a conceptual topic and as a career field (that is, she did little if any external, self-guided exploration of QIS).  By comparison, most of the Quantum Scholars students interviewed expressed not only an ``extra-curricular'' interest in exploring QIS but a strong desire to take ownership over the program and reinvent some of its implementations on their own terms.  

    %jamie not very interested in going into industry it doesn't sound like fun
    %jayden (engineer) expressed the most ambivalence to grad study but even then expressed plans for an "accelerated masters"
    
    %hillary accepted into grad school
    %bethany expressed plans to get masters in engineering after transfering to CU Denver and getting a bachelors and some experience (potentially in quantum)
    %charity transferring to cu boulder, thinking either MS or grad school after BS, CS (potentially quantum) but wants to get experience first
    %eleanor is meeting with people looking for jobs ``get some experience,'' expressed ambivalence about going to grad school without a clear awareness of jobs
    %mina wants to transfer to CU Boulder and get BS in CS, wants to work before going to grad school
    %eli transferring to Mines for CS
    
    %charity interested going to grad school, hillary actually accepted into a grad program. mina expressed potential interest two students, mina and %charity, continuing to the CU system from CCCS.

        %harvey: always had a decent grasp of what the public thinks of as quantum science
        %  - quantum computing class
        %dylan: quantum eng course at CU
        %timothy: senior year of HS and first semester at CU, talking with other students led to interest in QIS, quantum eng minor
        %jamie: summer 2022, did an internship at PNL
        %semaj: TED talk about string theory in 6th or 7th grade / real intro modern physics course at CU
        %jayden: started hearing about QM concepts in middle school, modern physics 
        %delaney: modern physics / quantum scholars

\subsection{Post-Participation Pathways and General STEM Affect}

Although all subjects interviewed expressed that they benefited from engagement with industry representatives and the greater quantum career field, as mentioned above the two program groups diverged in the ways in which they internalized that engagement.  We observed a continuation of this divergence in students' reported intended trajectories post-program.  While all six Quantum Scholars participants expressed a positive intention to continue into graduate study following the completion of their undergraduate degrees, the disposition toward immediate graduate study was much more varied and subdued for the QRX participants.  Among the Quantum Scholar students, only one, engineering student ``Jayden,'' expressed even a little ambivalence toward graduate study but still articulated a plan to complete an accelerated MS degree.  Physics student ``Jamie'' expressed a theme common to Quantum Scholars responses: ``I'm not very interested in going into industry; it just doesn't sound like fun.''  We note that even given a pronounced lack of interest in industry career pathways, multiple Quantum Scholars students expressed gratitude for the greater awareness of industry activity that the program provided them.

In constrast, the QRX students expressed a wide range of post-program plans inclusive of the consideration of graduate study but far more focused on gaining real-world experience, potentially in the quantum sector.  Two of the six interviewed students, Eleanor and another student ``Hillary,'' graduated with their bachelor's degrees at the same time as program completion.  Hillary reported that she had actually been accepted to graduate study at a university with a strong specialization in QIS and intended to complete a graduate degree in the field (this despite her relative lack of field career awareness, as described above).  Eleanor, by comparison, described using the industry connections described above to look for jobs, exhibiting a more circumspect disposition toward further schooling.  We note that Eleanor and Hillary were the only participants in the QRX program (that is, not just within the set of interview subjects) whose graduation coincided with the terminus of their program participation, and were two of three students from CU Denver, a four-year college, rather than a community college.  Of the other four students interviewed, three had been accepted for transfer from their TYC to a four-year college, and one was planning to do the same.  Three of those four students expressed at least tentative interest in graduate study but couched that interest behind a stronger desire to build their skillset and experience through employment (each explicitly mentioned potentially working in QIS) before pursuing further study.

Regardless of their program experience or career outlook, the vast majority of students across both programs expressed positive self-efficacy and a belief in their ability to succeed in their career trajectory.  11 of the 12 students interviewed expressed unequivocally that they felt they could succeed at their degree and at finding a job when the time came.  Only one QRX student, ``Mina,'' expressed explicitly that she did not feel confident in her ability to succeed in quantum science if she were to choose that pathway, mostly because the field was so unfamiliar to her even after program participation.  

Students in both programs were split on the impact that program participation had on their sense of self-efficacy in quantum.  Three QRX students and three Quantum Scholars students expressed that their confidence in their ability to succeed in QIS stemmed from their overall sense of self-confidence (i.e., a belief they could succeed at something they were interested in, regardless of field).  The other students expressed ``domain'' confidence for their degree field while acknowledging differences in their confidence for QIS prior to program participation -- however these students all reported that QRX participation positively impacted their confidence, either through exposure to QIS or through establishment of a supportive community.  Regardless of their motivations, goals, and career, and despite the difference in program design and objectives, it does appear that both programs have produced positive affective outcomes for their participant students.

It perhaps makes intuitive sense to arrive at the finding ``students in different programs have different experiences.'' However, we argue that the \textit{types} of differences evidenced by participants in the two programs may imply divergent goals and program needs for differing student populations; in the following section, we explore these divergences and their implications both for the different populations of students and for program designers looking to serve similar populations.  We will also discuss changes in program design prompted by these findings.

%%%%%%%%%%%%%%%%%%%%%%%%%%%%%%%%%%%%%%%%%%%%%%%%%%%%
\section{Discussion}
\label{sec:discussion}

\subsection{Thematic Differences Between Program Groups}

Unsurprisingly, students in the CU Boulder-based Quantum Scholars program tended to talk about their experiences in terms commensurate with their status as full-time students at a residential university, framing program activities and opportunities as educational and relational enrichment along their path toward a degree and probable graduate study.  Lectures from quantum industry leaders (even CEOs of local startups) and opportunities for networking with industry members were viewed as enjoyable opportunities to expand horizons and to develop and broaden their understanding of the quantum ecosystem, but students did not report a major shift in their career goals or expectations for post-program quantum engagement.  QRX students, on the other hand, reported not only greater appreciation specifically for the direct connection to industry members but appreciable shifts in their consideration of the quantum industry as a career field.  What factors influence this perceived divergence?  Of course, one simple explanation for the deeper appreciation for industry relations and career support over academic enrichment in the QRX student group is that the program was designed with these goals in mind and, as a consequence, features these opportunities more explicitly.  

However, we note some differences in the dispositions or pre-participation goals of participant students in the two groups that may also be contributing factors to their varying perspectives on program participation.  First, QRX students were more likely to view their schooling as a means to a vocational end rather than as ``education for its own sake.''  Even the students who joined QRX from four-year CU Denver expressed a more pragmatic, job-focused disposition toward their education and toward program participation than the CU Boulder-based Quantum Scholars students, who tended to highlight the academic/cerebral nature of their schooling.  Quantum Scholar students also tended to describe elements of their pre-collegiate experiences more aligned with the traditional nature of their current schooling: greater levels of self-guided scientific exploration, including of QIS; broader awareness of the pathways available to them in and following college; greater reflection on fields of interest and potential specialization; etc.  By comparison, QRX students were more likely to report non-traditional backgrounds (e.g., working before college) or to express a pragmatic motivation for enrollment at their particular school; for example, multiple students expressed that their decision to attend their two-year college had been predicated upon consideration of expenses after immigration to the US for the purposes of schooling.

Taken together, these findings highlight key differences in the way that different groups of students may engage in higher education -- we note that, even with its relatively stronger focus on enrichment from the start (scholarships vs internships, student-led design elements similar to ``clubs,'' etc.) vs. QRX's focus on professional development (resume-building workshops, industry mixers), QScholars students still framed their desires for programmatic changes or improvements in terms of supporting the broadening of their largely academic experiences.  This makes sense given their relatively traditional set of backgrounds and professed trajectories; for students heavily ensconced in the culture of academia, engagement with QIS is perhaps more likely to look like engagement with QIS topical specialization.  For QRX students, school was described more frequently as a means to acquire skills and credentials needed for employment.  An extracurricular program focused on exploring the nuances of quantum theory would almost certainly not be as appropriate or as valued in this context as a program providing opportunities to strengthen resum\'es, network, etc.  We emphasize that these differences in disposition do not indicate that one or the other of these student groups are smarter, more capable, more ``fit'' for a given mode of higher education engagement, etc.; rather, through articulating these divergent dispositions, we argue that students are providing insight into their motivations, needs, and goals -- crucial elements to be aware of for program designers in these spaces.

\subsection{Two interesting profiles: Hillary and Jayden}

To what extent are the dispositions reflected in student responses ``typical'' of their respective populations?  Or, put another way, where are the boundaries between the more ``pragmatic'' and ``academic'' dispositions observed in our student groups that might influence a broader or more focused program design?  To help answer this question, we highlight two particular student profiles that may serve as ``edge cases'' for their respective student groups, QRX student Hillary and Quantum Scholars student Jayden, both mentioned above in Section \ref{sec:results}.

Hillary is a QRX student who expressed that she had not encountered QIS as a field at all before program participation, i.e. program participation was her first exposure to QIS.  She first joined QRX as a junior at her four-year institution (CU Denver, which is classified as an R1 but does not have physics graduate study like CU Boulder and is a much smaller department; is also hispanic-serving institute), then participated for two years.  Hillary reported that she had been accepted for graduate study at a strong QIS-focused graduate program, and explicitly named QRX as the impetus for her deeper interest in the field.  

Jayden is Quantum Scholars student with an electrical and computer engineering (ECEE) major (the rest of the students we interviewed were physics or engineering physics).  He, out of all the interviewed students, expressed the strongest ``industry-focused'' career goals and interests, and the strongest awareness of how industry functions (often articulating an understanding of how government funding in the field has led to the current rush for funding of industry players, the importance of startup funds for a new industry, etc.).  Where the other students often expressed a desire for \textit{more} subject-focused lectures and topical explorations, Jayden actually expressed disappointment that most of the speakers, despite being in industry, came from a physics background and were talking more about the AMO elements of quantum rather than the engineering/technical components of the field.  Jayden plans to do an accelerated master's program following his graduation.

Again, we highlight these students as ``edge'' cases in their respective programs in terms of their described career pathways, expressed interests in quantum science, and post-graduation plans.  Most QRX students, if they discussed graduate study at all, did so in the context of additional training for employment, and often following a span of time gaining ``real experience'' in the career field.  Hillary is the only student who expressed even a desire to move directly from undergraduate study to grad school, much less an actual enacted plan for it.  For Quantum Scholars, Jayden exhibited an awareness of the ways in which industry and academic fields coordinate and communicate in quantum science that we did not observe in the other interviewed students, and expressed a desire to work in fabrication, electronics-building, or a similar, more applied subfield after completing his master's degree.  Both students seem to bridge the gap between what we might describe as the ``theoretical'' and ``applied'' directions of their respective programs, and it is perhaps easier to imagine these two students thriving in the other program than their peers.

In both cases, however, the student still expressed views of their respective program commensurate with their peers (i.e., Jayden still framed Quantum Scholars largely as enrichment and Hillary still framed QRX largely as career preparation).  We highlight these two students to bring attention to the fact that even though their disposition toward participation in the quantum industry may be similar, the two students' outlooks on program participation still reflect their differing contexts.  It appears as though, even for students with dispositions or objectives that bridge the contextual gaps between the two groups of program participants, program design must still attend to issues of background, student needs, and framing of participation.  We thus turn to questions of program design and discussion of the ways in which these findings can influence (and have influenced) adaptations in the ways these programs serve their student populations.

\subsection{Considerations for Program Design}

First, we note that even this small dataset supports a finding reported by Rosenberg \textit{et al.,} namely that students largely do not understand the quantum science field or how they may fit into it \cite{rosenberg2024science}.  Across both groups, regardless of academic level or disposition toward the field, students reported only a shaky understanding of the quantum career landscape, and expressed varying degrees of proficiency even articulating a conceptual understanding of quantum science.  Similar to Ref. \cite{rosenberg2024science}, however, we note that students generally expressed strong interest in quantum careers despite their lack of knowledge; in particular, as described in Section \ref{sec:results}, QRX students went from being largely unaware of QIS entirely to considering it as a possible career path.  

How should these considerations inform program development?  For programs like QRX, where facilitation of pathways to quantum careers is an explicit goal, it may be beneficial to create more opportunities to gain a ``foundational'' quantum understanding.  Note that, as discussed in Section \ref{sec:intro}, this does not necessarily equate to a specialist understanding; for ``quantum-conversant'' workers, a broad conceptual awareness is likely to be sufficient, as deep knowledge of quantum topics is less important than the technical and lab skills required of the position \cite{fox_preparing_2020}.  For programs like Quantum Scholars, where students' interests may be more academic and less job-focused, opportunities to explore quantum topics and interact with subject matter experts may be more important and relevant than opportunities to network or learn about the quantum industry.  

Even with a difference in preference for the framing and content of events connecting student groups to quantum professionals (e.g., NIST tour \textit{vs.} industry mixer), students across the board in both groups articulated a strong appreciation for events that put them face-to-face with professionals in the quantum field.  Similarly, both groups expressed a strong appreciation for the community they built in the program, even if, again, the nature of those communities differed according to the culture of the program.  For a largely residential undergraduate university such as CU Boulder, it makes sense that students were more likely to report community benefits like finding a roommate, making a new friend, and so on; for QRX students in their disparate collegiate settings, it similarly makes sense that students reported appreciating the opportunity to make ``LinkedIn friends'' with other program participants and industry members.  In designing workforce programs that appeal to the social and professional networking needs and interests of students, we thus implore educators to consider that, while many do, not every student frames extracurriculars like these programs as interest-based ``clubs.''  Designing programs with the social goals of students in mind is crucial to ensure their success.

Finally, we touch upon the consideration of participant affect in program design.  It is well-understood by now that persistence in STEM is linked to affective constructs such as belonging, self-efficacy, STEM identity development, etc. \cite{lewis2016fitting,hyater2018critical,kinzie2008promoting,lock2013physics,lane2016research}.  As discussed above, strong self-efficacy beliefs were not observed to be more prevalent among either group of students; both groups contained students that expressed that participation in their program had helped improved their belief in their ability to succeed at quantum-related pursuits.  Of course, design of educational programs should always attend to issues of affect, regardless of context.  Whether at residential schools, high-performing R1s, trade-focused community colleges, primarily undergraduate institutions, etc. -- whatever the academic context, students benefit from explicit programming that facilitates the belief that they can succeed at their efforts in the domain at hand.  We note as well, however, that for underrepresented students especially, affect-focused design philosophy can be married to certain program design elements to improve outcomes.  Cohort-based models, increased opportunities to explore career fields, and other methods can all be used to facilitate persistence and development of STEM identity \cite{narayanan2018upward,blair2012impact,toven2015increasing}.  Both programs use a cohort model and incorporate elements of student ownership over program design and implementation; we argue that this focus on centering student community and agency has played a large role in the fact that students in both groups relatively universally expressed both a positive community experience and positive affect after participation.

%%{\color{red} HIGH-IMPACT PRACTICES GO HERE?}

\subsection{Implementation of Design Feedback: How Have These Programs Changed?}

We close the discussion by describing briefly some of the ways in which both programs have incorporated student feedback in their design and implementation.  We highlight increasing divergences in the ways the programs are implemented and tie those divergences to the needs and interests of students.

\subsubsection{QRX Adaptations}

For the multi-site QRX program, the key pieces of critical feedback received from participants were (1) the desire for more in-person events, (2) the desire for more explicit quantum foundational content, and (3) the desire for more opportunities to connect to industry members.  We note that all three of these pieces of feedback are criticisms of ``quantity,'' not of quality or degree; taking them into account, along with feedback on program community and culture, program design for QRX is shifting to more fully center opportunities for students to develop concrete professional skills and strong confidence as ``quantum-conversant'' future workers.  At the time of publication, QRX is between cohorts; a number of changes are being considered for implementation in future cohorts.  One such change would greatly increase  the number and variety of in-person networking opportunities and front-load those events in the first half of the program (e.g., industry Q\&A panels building from students' questions coming out of an early industry mixer).  Another change is to present quantum topical opportunities more as skill development workshops and less as ``mini courses'' -- even to the extent of considering providing micro-credential opportunities for students engaging with particular quantum platforms such as Ocqtang or CUDAQuantum \cite{tingle2024oqtant,kim2023cuda}.  Finally, the program is also working more closely with quantum industry leaders in the Boulder-Denver to build a strong culture of student-industry engagement at the non-R1 level.  

The impact of changes like these remains to be seen but assessment strategies are already in place for future cohorts.  We note that, crucially, we are leveraging Q-SEnSE's larger administrative and logistical resources to ensure that program participants are not required to create or implement these changes themselves;  QRX as a program is designed to provide opportunities for students to plug in to quantum education and workforce preparation when their interests, time, and life constraints allow.  In some sense, a healthy instantiaion of QRX may come to look less like an extracurricular activity and more like a professional society as students take greater ownership over their path to industry.

\subsubsection{Quantum Scholars Adaptations}

Quantum Scholars has also adapted to student feedback; as it began a year before QRX, it has been able to implement a number of changes in its second year of major operation.  As with QRX, the program has increased the number of social events throughout the year -- in fact, the number of meetings has doubled from once to twice a month, with half of the overall meetings now an informal discussion on quantum subjects led by students over a meal.  The program has also increased the number of tours, similar to the above-mentioned NIST tour, that broaden awareness of non-academic quantum activity and span the space between academic and industry career fields.  In order to encourage increased and broader student participation, the program has also increased the amount of scholarship support for applicants.

The major design-based change in the Quantum Scholars program is a dramatic increase in student ownership over the program and student leadership in its design and implementation.  Largely via the initiative of inaugural cohort members, the program has expanded to include more peer mentoring, and has established a student ``council'' that represents student interests within the program.  In its second year, Quantum Scholars also implemented a hackathon (as described above in Section \ref{sec:context-qscholars}) -- this effort was not only proposed by two Quantum Scholars students, it was designed and implemented almost entirely via student effort.  Students designed the event, solicited support from experienced graduate student mentors to facilitate quantum computing problems, and presided over the event largely by themselves; academic faculty advisors and the department provided resources and logistical support, but otherwise left the students to their own devices.  This greater focus on student ownership of programmatic design is perhaps the strongest point of deviation between the two programs; the context of the Quantum Scholars program strongly supports the integration of the program into the everyday lives of its participants in a way that the context of QRX simply doesn't, a strength of the program for the students it purports to serve.

%%%%%%%%%%%%%%%%%%%%%%%%%%%%%%%%%%%%%%%%%%%%%%%%%%%%
\section{Conclusions \& Future Work (draft)}
\label{sec:conclusions}

First, we celebrate the fact that, with diverse student needs, diverging goals and interests, and unique backgrounds, it appears to be the case that both programs were successful in their implementation.  Student feedback for both programs indicates that a variety of design techniques can have a positive impact on students' dispositions toward QIS and their interest in QIS careers.  Both programs have incorporated negative feedback and are well-positioned to continue to grow and serve their respective student population groups.  Thus, we emphasize a key conclusion from these data: quantum education and workforce development is not an ``either/or'' proposition.  There is space -- indeed, we argue that there \textit{must} be space -- in quantum education for content aimed at promoting a specialist education in an academic setting and for content aimed at attaining only a basic quantum understanding but providing a richer vocational foundation.  We encourage quantum educators, particularly those at R1s, to consider ways in which their efforts may assume a ``content first'' mindset or may implicitly assign greater value to the pursuit of quantum specialization over practical quantum engagement.  

Multiple authors of this manuscript have expertise in physics education in non-classroom settings, in programs similar to these or more explicitly ``informal'' in nature; often in such programs, key articulated goals are ``engagement'' and ``exposure'' over content learning.  Even in the case of programs like QRX and Quantum Scholars that contain out-of-classroom elements and focuses on exposure to quantum ideas, we encourage educators to ask the question, ``exposure to what?'' Exposing students to guest lectures from quantum experts is, naturally, a very different design and implementation question than exposing them to practical opportunities to apply for entry-level quantum jobs.  Again, we exhort designers and educators to consider the needs and objectives of the populations they purport to serve.  This issue is more than simply an academic consideration; the QIS field is widely agreed that quantum-ready workers are needed in a \textit{variety} of positions, subfields, and expertise levels.  The major goal of QRX as a program is to ensure that students who might thrive in QIS but may lack the prior structural and community resources of a student in Quantum Scholars can still participate in the field if they so choose.  We maintain that these programs, or programs like them, can learn from each others' designs -- indeed, we are already planning opportunities for students to ``cross-pollinate,'' especially given that increasing numbers QRX students may find themselves at CU Boulder with the potential to participate in Quantum Scholars in the future as they transfer out of their two-year college and into a four-year setting.

As the quantum education and workforce conversation grows, it is natural to talk about ``best practices'' in terms of designing courses and programs that serve students.  One crucial point suggested by these findings is that, depending upon needs and goals of the student population in question, there may be no single set of best practices for answering the question of how to populate the quantum workforce.  Rather, we argue for a quantum education ethos informed by research-based \textit{best principles} whose practical application changes according to context.  Crucially, any approach that purports to attend to ``best practices'' for its target population must necessarily contend with DEI considerations for its population, to ensure that program design goals actually align with the expressed goals, interests, and needs of target populations.  We urge quantum workforce programs to treat their participants -- whether in an R1 setting, a community college setting, or a K12 setting -- as co-creators and invested parties, rather than as audience members.

This study makes some salient suggestions about the needs of different student populations, but we acknowledge that its data set is small, if rich.  In order to fully understand the contexts of the diverse groups of future quantum workers and to make generalizable claims about beneficial practices for those groups, we encourage quantum education researchers to consider large-scale quantitative studies not just of the classrooms and university spaces where QIS is taught, but the workforce development spaces where workers are reached.  We anticipate continuing research on these programs ourselves as they continue to grow, but note that our context is likely very different from the context for programming implemented in another quantum-rich locale or a part of the country with little quantum industry presence.

As both quantum education and workforce development programs are created and implemented, there is a real opportunity for the field to create a truly equitable and representative quantum workforce.  However, in order to achieve this goal, educators should act \textit{now}, as the field spins up.  To borrow language from the Universal Design for Learning (UDL) movement, it is far easier and more effective to design for equity in quantum education than it will be to ``retrofit'' once the field is fully formed \cite{schreffler2019universal}.  As more and more research provides a better understanding on the impacts of quantum education efforts, we urge researchers to build on this study and investigate ways in which marginalized populations already do and can be encouraged to participate in QIS -- especially if we as educators, employers, and workforce leaders claim to desire their participation in the quantum workforce.

\begin{acknowledgments}
The authors would like to gratefully acknowledge the participation of students in both the Quantum Scholars and QRX programs, as well as the program partners, both academic and industry, that enable these programs to serve the needs of their student participants, including the CU Boulder Department of Physics and the College of Engineering and Applied Sciences, the Colorado Community College System, the CU Denver Department of Physics, and the CUBit quantum initiative.  This work is supported by the Q-SEnSE NSF Quantum Leap Challenge Institute (NSF QLCI OMA-2016244).  Any opinions, findings, and conclusions or recommendations expressed in this material are those of the author(s) and do not necessarily reflect the views of the National Science Foundation.
\end{acknowledgments}


%%\bibliography{zotero_bib}% Produces the bibliography via BibTeX.
%% ATTN: whatever bib file we use here, ensure that the entries work for BibTeX

\end{document}